\documentclass[aps,showpacs,prb,superscriptaddress,floatfix,twocolumn,citeautoscript]{revtex4-2}
\usepackage{graphicx}
\usepackage{amsmath}
\usepackage{amssymb}
\usepackage{float}
\usepackage{bm}
\usepackage{dcolumn}
\usepackage{subfigure} 
\usepackage{hyperref}
\usepackage[usenames]{color}
\usepackage{multirow}
\usepackage{subfigure}
\usepackage{lipsum}

\hypersetup{pdfborder=0 0 0,colorlinks=true,citecolor=blue,linkcolor=blue}
\input epsf
\DeclareGraphicsExtensions{.pdf,.jpeg,.png, .eps}

\bibliography{refs} 
\begin{document}
	\title{Strong anisotropic optical properties of 8-Pmmn borophene: a many-body perturbation study}
	
	\author{N. Deily Nazar}
	\affiliation{Department of Physics, Shahid Beheshti University, G. C., Evin, Tehran 1983969411, Iran}
	\affiliation{Department of Physics, University of Antwerp, Groenenborgerlaan 171, B-2020 Antwerpen, Belgium}
	\author{T. Vazifehshenas}
	\email{t-vazifeh@sbu.ac.ir}
	\affiliation{Department of Physics, Shahid Beheshti University, G. C., Evin, Tehran 1983969411, Iran}
	\author{M.R. Ebrahimi}
	\affiliation{Department of Physics, Shahid Beheshti University, G. C., Evin, Tehran 1983969411, Iran}
     \author{F.M. Peeters}
     \affiliation{Department of Physics, University of Antwerp, Groenenborgerlaan 171, B-2020 Antwerpen, Belgium}
 
  ]
	\begin{abstract}
		
Using first-principle many-body perturbation theory, we investigate the optical properties of 8- borophene at two levels of approximations; the $\bf GW$ method considering only the electron-electron interaction and the $\bf GW$ in combination with the Bethe-Salpeter equation including electron-hole coupling. The band structure exhibits anisotropic Dirac cones with semimetallic character. The optical absorption spectra are obtained for different light polarizations and we predict strong optical absorbance anisotropy. The absorption peaks undergo a global redshift when the electron-hole interaction is taken into account due to the formation of bound excitons which have an anisotropic excitonic wave function. 

\end{abstract}


\maketitle

\section{Introduction}

The emergence of two-dimensional (2D) Dirac materials exhibiting novel and fascinating physical properties, has led to extensive theoretical and experimental studies and many fundamental research breakthroughs in recent decades\cite{T. Wehling,E. Fradkin,N.P.Armitage}.
Graphene a nanosheet of carbon atoms with an isotropic Dirac band dispersion and tunable electrical, thermal and optical properties is the most famous in this category which e.g. can be obtained by mechanical exfoliation of graphite \cite{K. S. Novoselov}.
Similar to graphene, other 2D Dirac materials like silicene\cite{P. Vogt}, germanene\cite{M. E. Davila} and stanene\cite{F. Zhu} are potential candidates for various nanotechnology applications.  
Recently, a new elemental 2D Dirac material formed by boron atoms, called 8-Pmmn borophene, has been predicted. This is one of the most stable polymorphs of 2D boron sheets which has a linear low-energy dispersion like graphene, but with tilted and anisotropic Dirac cones\cite{X. F. Zhou,Z. Q Wang}. In orthorhombic 8-Pmmn borophene, the conduction and valence bands touch at the Dirac point in the Brillouin zone indicating its semimetallic nature.

\begin{figure} [ht!]
	\includegraphics [width=0.85\linewidth]{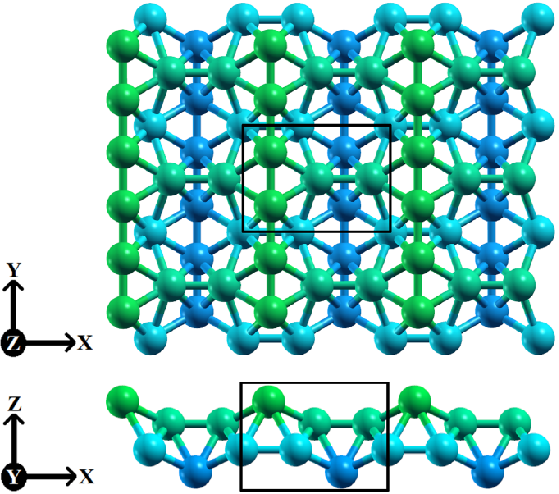} %
	\centering 
	\caption{Top and side views of the optimized 8-Pmmn borophene sheet exhibiting
		a buckled crystal
		structure formed by two sublattices. The solid rectangles indicate the unit cell.}
	
	\centering
	\label{fig:borophene}
\end{figure}
First studies on the electronic structure of 8-Pmmn borophene was performed by using first-principles calculations within the Perdew-Burke-Ernzerhof (PBE) generalized gradient approximation (GGA) functional for the exchange-correlation and the projector-augmented-wave method as implemented in VASP\cite {X. F. Zhou,A. Lopez-Bezanilla}. It was found that the structure of 8-Pmmn is composed of hexagonal and chain motifs. Also, based on a Born effective charge analysis, it was concluded that 8-Pmmn borophene contains two types of nonequivalent boron atoms with opposite effective charges which identifies it as the first ionic elemental monolayered structure.

Fig.\hspace{0.1 cm}\ref{fig:borophene} shows top and side views of the crystal structure of 8-Pmmn borophene. In order to show the buckled structure of the hexagonal sublattice clearly, different colors for the boron atoms are used along the $z$ direction. DFT calculations have predicted that external stress\cite {A. Lopez-Bezanilla}, normal incidence of electromagnetic radiation \cite {A. E. Champo} and hydrogen adsorption \cite{Z. Q Wang} can induce a transition from semimetallic to semiconducting behavior in this allotrope of 2D boron. The magnetotransport \cite{S. F. Islam}, mechanical\cite{J. Yuan}, and thermal\cite {P. Sengupta} properties of 8-Pmmn borophene have been investigated and effects due to the tilted anisotropic Dirac cones were found. In another work\cite{S. Verma}, the anisotropic Drude weight and optical conductivity of this phase of borophene have been studied in detail. In a DFT study\cite {suman}, the electronic and optical properties of 8-Pmmn borophene doped with Li, Be, C and H atoms were investigated. A transverse optical conductivity was predicted\cite{A. Singh}. Moreover, the adsorption of transition metal adatoms (Fe, Co and Ni) on 8-Pmmn borophene has been explored and it has been shown that Co and Fe convert the adatom/8-Pmmn borophene system into a ferromagnetic one\cite{S. Tomar} which makes it potentially interesting for spintronics applications. 

Many-body interactions including electron-electron (e-e) and electron-hole (e-h) are of great importance in low-dimensional systems due to the strong quantum confinement effects and small Coulomb screening\cite{A. L. Fetter,L. Yang,W. Wei and}. In 2D materials, many-body calculations have been widely used to characterize the electronic, electrical and optical properties of such structures\cite{M. Yarmohammadi,V. Tran,W. Wei and}. 
Screening and collective electronic excitations (plasmons) arise from the long-range Coulomb interaction between the electrons which have been studied in 8-Pmmn borophene\cite{numb16,numb15}. In the context of DFT, the optical properties of graphene, silicene and germanene have been calculated where many-body effects are taken into account by using the non-local hybrid functional \cite{L. Matthes} or by the Green's function perturbation theory, i.e., $GW$ plus Bethe Salpeter equation ($GW-BSE$) \cite{ L. Yang, W. Wei}. Higher quasi-particle energies and larger overlap between electron and hole wave functions were achieved as a result of the reduced screening effect and the reduced dimensionality of such systems with respect to their bulk counterparts \cite{W. Wei and}. The optical absorption peaks are red shifted in graphene, silicene and germanene due to strong excitonic effects in good agreement with experiment \cite{L. Yang}.

Motivated by the interesting excitonic features of 2D Dirac materials with isotropic energy dispersion, we investigate here the optical properties of 8-Pmmn borophene including both many-body e-e and e-h interactions. First, we calculate $GW$ band structure of 8-Pmmn borophene which is expected to provide a more realistic description of its electronic structure compared to previous DFT-LDA calculations\cite{X. F. Zhou,A. Lopez-Bezanilla}. Employing the $GW-BSE$ and $GW-RPA$ approaches, we compute the optical excitation spectra of 8-Pmmn borophene with and without e-h interaction and show that its optical properties are highly anisotropic. We also obtain the excitonic wave functions for two different excitons associated with two important absorption peaks when the radiation polarization vectors are along $x$ and $y$ directions. Since our calculations is based on many-body perturbation theory, the effect of external fields on the electron energy spectrum is very small, so the predicted additional asymmetry induced by a linearly polarized strong electromagnetic field\cite{kristinsson} on the Dirac cones of 8-Pmmn borophene is negligible in our work.

The outline of this paper is as follows. In the next section, we describe the formalism for calculating the optical absorption spectra. Then, we present and discuss our results in detail. Finally, the last section is devoted to the conclusion.

\section{Theory and computational details}  
The low energy electronic band structure of 8-Pmmn borophene exhibits a tilted-anisotropic linear dispersion along the high symmetry line connecting the $\Gamma$ and $Y$ points. First, we perform $ab \hspace{0.1 cm} initio$ DFT calculations within the local density approximation (LDA) for the exchange and correlation energy functional to compute the band structure of 8-Pmmn  . The norm-conserving pseudopotentials are used to include the electron-ionic core interactions. A plane-wave (charge density) cutoff energy of ${100 \hspace{0.1 cm} (400 )Ry}$ is taken and a ${24\times{34}\times{1}}$ Monkhorst-Pack $k$-point grid is used for the Brillouin zone sampling. A vacuum distance of 27{\AA} along the ${z}$ direction is included to prevent spurious interactions. Relaxation of the lattice parameters and atomic positions are carried out in order to reduce the forces on the atoms below ${0.005}$ eV/\AA. Our DFT calculations are done using Quantum ESPRESSO\cite{P. Giannozzi,P.G}. As mentioned earlier, DFT-LDA is not a reliable method for describing the excited state properties of materials. It is well-known that the $GW$ method \cite{F. Aryasetiawan} provides a sufficiently accurate modification to the DFT eigenenergies by introducing the self-energy correction, $\Sigma$. In this approach, the electron self-energy is expressed in terms of the single-particle Green's function, $G$, and the screened potential, $W$. 
The starting point is the Dyson equation for the Green’s function

\begin{equation}
	\begin{aligned}
		G=G_{0}+G_{0}\Delta\Sigma G \,
		\label{eq:Green}
	\end{aligned}
\end{equation}
which can be solved perturbatively to obtain the quasi-particle energies and thus, the $GW$ electronic band structure. Here, $G$ is the exact Green’s function of the interacting system, $G_{0}$ is the Green’s function calculated from the DFT eigenenergies and eigenfunctions and $\Delta\Sigma=\Sigma-V_{xc}$ is the difference between the self-energy and exchange-correlation potential. In our calculations, we use a non self-consistent solution of the Dyson equation for the self-energy, i.e. the $G_{0}W_{0}$ approach, in which the dielectric function of the screened e-e potential is calculated from the generalized plasmon-pole approximation. We used 72 unoccupied states (6 times the number of occupied states) to achieve a converged dielectric function.  Considering the e-h interaction as well as e-e interaction and constructing a two-particle Green's function, it is possible to compute the photo-excited states and optical absorption spectra within the $GW-BSE$ approach\cite{G. Onida,M. Rohlfing}. By expanding the excitonic wave functions in terms of the $GW$ quasi-particle wave functions, the $BSE$ Hamiltonian turns into a two-particle eigenvalue problem:  	
\begin{equation}
	\begin{aligned}
		\sum_{{\nu} {c} {\mathbf{k}}}\sum_{\nu^{'} c^{'}\mathbf{k^{'}}}<c\nu \mathbf{k}|K_{e-h}|c^{'}\nu^{'} \mathbf{k}^{'}>\mathbf{A}_{c^{'}\nu^{'} \mathbf{k}^{'}}^{\lambda}\\
		\	+(E_{c\mathbf{k}}-E_{\nu \mathbf{k}})\mathbf{A}_{c\nu \mathbf{k}}^{\lambda}   =\mathbf{\mathrm{E}_{\lambda}}\mathbf{A}_{c\nu \mathbf{k}} \,
		\label{eq:hamiltoni}
	\end{aligned}
\end{equation}

\noindent where $|c\nu\mathbf{k}>$ indicates the pair of quasi-electron and quasi-hole states, $|c\mathbf{k}>$ and $|\nu\mathbf{k}>$. ${\mathrm{E}_{\lambda}}$ and $A^{\lambda}_{cv\mathbf{k}}$ are the eigenvalues (i.e. excitonic energies) and excitonic eigenvectors, respectively. $E_{c \mathbf{k}}$ ($E_{\nu \mathbf{k}}$) denotes the $GW$ quasi-particle energy in the conduction (valence) band. $K_{e-h}=\bar{W}-2\bar{V}$ is the $BSE$ kernel with $\bar{W}$ accounting for the screened Coulomb interaction between electrons and holes  \cite{numb15} 

\begin{equation}
	\begin{aligned}
		\bar{W}(c\nu\mathbf{k},c^{'}\nu^{'}\mathbf{k_{1}})=\dfrac{1}{\Omega}\sum_{\mathbf{G}\mathbf{G^{'}}}
		v(\mathbf{q}+\mathbf{G^{'}})\epsilon_{\mathbf{G}\mathbf{G^{'}}}^{-1} \\ 
		\times <\nu^{'}\mathbf{k_{1}}|e^{-i(\mathbf{q}+\mathbf{G^{'}}).\mathbf{r}}|\nu\mathbf{k}> 
		<c\mathbf{k}|e^{i(\mathbf{q}+\mathbf{G}).\mathbf{r}}|c^{'}\mathbf{k_{1}}>
		\delta_{\mathbf{q},\mathbf{k-k_{1}}}
	\end{aligned}
\end{equation}
and $\bar{V}$ is the bare e-h exchange interaction

\begin{equation}
	\begin{aligned}
		\bar{V}(c\nu\mathbf{k},c^{'}\nu^{'}\mathbf{k_{1}})\\=\dfrac{1}{\Omega}\sum_{\mathbf{G}\neq 0}\nu(\mathbf{G})<\nu^{'}\mathbf{k_{1}}|e^{-i\mathbf{G}.\mathbf{r}}|c^{'}\mathbf{k_{1}}><c\mathbf{k}|e^{i\mathbf{G}.\mathbf{r}}|\nu\mathbf{k}>.
	\end{aligned}
\end{equation}

Here, ${\Omega}$ is the unit cell volume, $v(\mathbf{G})$ is the bare potential and $\epsilon_{\mathbf{G}\mathbf{G^{'}}}^{-1}$ represents the inverse of the microscopic dielectric screening matrix which is related to the macroscopic dielectric function, $\epsilon_{M}(\omega)$, as follows\cite{A. Marini}:

\begin{equation}
	\begin{aligned}
		\epsilon_{M}(\omega)=\dfrac{1} {\mathbf\epsilon_{\mathbf{G=0}\mathbf{G^{'}=0}}^{-1}(\mathbf{q}=0,\omega)}.
	\end{aligned}
\end{equation}

In the $BSE$ formalism, the macroscopic dielectric function which is a key quantity whose imaginary part directly determines the optical absorption spectrum, is given by:      	      	 

\begin{equation}
	\begin{aligned}
		\epsilon_{M}(\omega)=1-\lim\limits_{q\rightarrow{0}}\dfrac{8\pi}{\rvert{q^2}\rvert\Omega}
		\sum_{vc\mathbf{k}}\sum_{\nu^{'}c^{'}\mathbf{k^{'}}}<\nu\mathbf{k}-\mathbf{q}
		|e^{-i\mathbf{q}.\mathbf{r}}|c\mathbf{k}>\\
		\times<c^{'}\mathbf{k^{'}}|e^{i\mathbf{q}.\mathbf{r}}|
		\nu^{'}\mathbf{k^{'}}-\mathbf{q}>
		\sum_{\lambda}\dfrac{\mathbf{A^{\lambda}_{c\nu\mathbf{k}}}
			(\mathbf{A^{\lambda}_{c^{'}\nu^{'}\mathbf{k^{'}}}})^*}{\omega-\mathrm{E}_{\lambda}}.	
	\end{aligned}
\end{equation} 
Upon diagonalizing the $BSE$ equation (Eq. (2)), the excitonic energies and wave functions are obtained and used for constructing $\epsilon_{M}(\omega)$. Our excited state calculations are done using the YAMBO code\cite{D. Sangalli,A. Marini}. In $GW-BSE$ calculations, a truncated Coulomb potential along the non-periodic direction is used.

\section{Results and discussion}
\subsection{The electronic properties}

\begin{figure} [ht!]
	\includegraphics [width=1.1\linewidth]{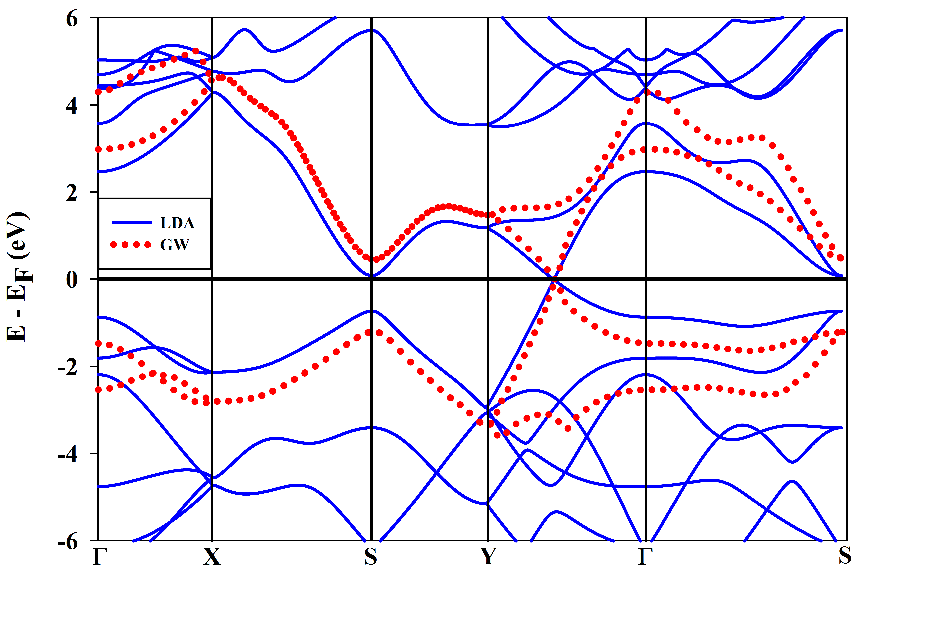}  
	\caption{The DFT-LDA electronic (solid blue lines) and $GW$ quasi-particle (dotted red lines) band structures of 8-Pmmn borophene.}
	\centering
	\label{fig:Band_eps}
\end{figure}

In Fig.\hspace{0.1 cm}\ref{fig:Band_eps} we show the electronic band structure of the free-standing intrinsic 8-Pmmn borophene which is calculated using the DFT method within the LDA approximation. To improve the accuracy of the results, the $GW$ calculations including the quasi-particle self-energy correction has been performed and the resulting energy bands are shown in this figure, as well. It can be seen that the $GW$ calculations confirm the existence of the anisotropic Dirac cone at the Fermi energy, $E_{F}$, and the semimetal behavior of 8-Pmmn borophene as previously predicted by other DFT studies\cite{X. F. Zhou,A. Lopez-Bezanilla}. Actually, the anisotropic Dirac cone band structure of 8-Pmmn borophene is attributed to the mirror symmetry in the vertical direction of the crystal lattice\cite{Guido van Miert,Xuming Qin}and GW
approximation enhances just e-e interaction, therefore the cone characteristics like anisotropy is preserved in both the LDA and GW approximations.

\begin{figure} [ht!]
	\includegraphics [width=1.1\linewidth]{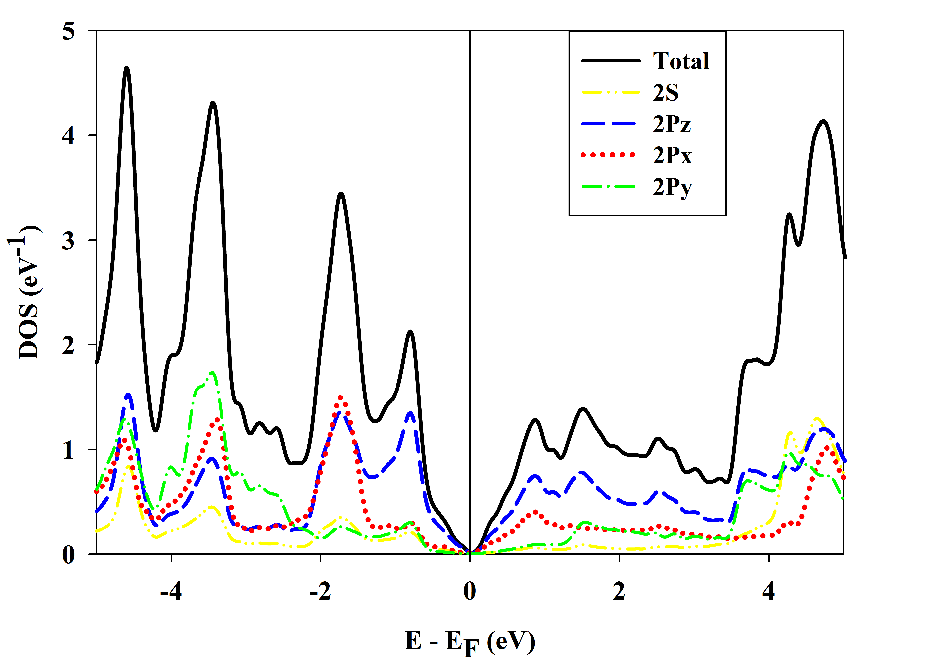}  
	\caption{The total and projected density of states of 8-Pmmn borophene calculated using DFT-LDA method.}
	\label{fig:PDOS_NEW_1}
\end{figure}

Many-body effects increase the bandgap at the $S$ point from $0.81$ eV(LDA) to $1.67$ eV(GW) and at the $\Gamma$ point from $3.34$ eV(LDA) to $4.58$ eV(GW). In addition, it is found that the $GW$ corrections result in a significant Fermi velocity renormalization near the Dirac point, similar as previously found for the cases of intrinsic graphene\cite{C. Attacalite} and buckled silicene\cite{S. Huang}.
We have obtained $v_{F}=0.49\times{10^{6}} $m/s (LDA) and $v_{F}=0.67\times{10^{6}} $ m/s (GW) in $Y$-direction and $v_{F}=1.12\times{10^{6}} $ m/s (LDA) and $v_{F}=1.51\times{10^{6}} $ m/s (GW) in the $\Gamma$-direction. Notice that the Fermi velocity enhancement is large in 8-Pmmn borophene which can be attributed to a weaker e-e screening effect in this 2D Dirac material as compared to graphene and silicene.

Also, the total and projected density of states (DOS) of 8-Pmmn borophene are calculated and displayed in Fig.\hspace{0.1 cm}\ref{fig:PDOS_NEW_1}. Notice that the contribution from $p_{z}$ orbitals is dominant in the vicinity of the Dirac point, in agreement with Ref.\cite{A. Lopez-Bezanilla}.          

\subsection{The optical absorption spectrum}

\begin{figure} [ht!]
	\includegraphics [width=1.10\linewidth]{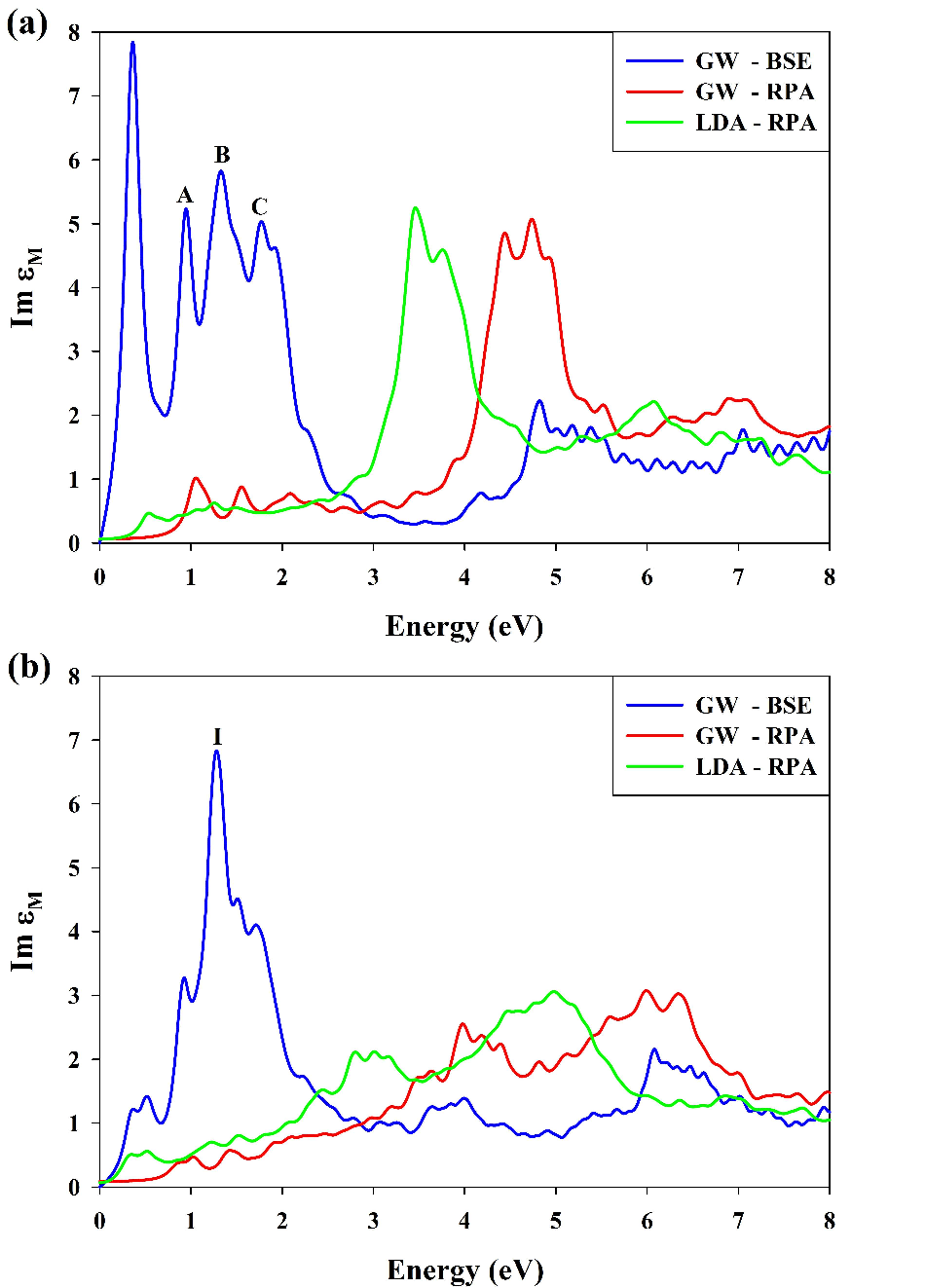}  
	\centering      
	\caption{The absorption spectra of 8-Pmmn borophene along (a) $y$ and (b) $x$ directions calculated using the $LDA-RPA$, $GW-RPA$ and $GW-BSE$ methods.}
	\label{fig:absorbtion} 
\end{figure}

In order to calculate the optical absorption spectrum, we go beyond the one-particle Green's function formalism by taking into account the electron-hole two-particle Hamiltonian extracted from the $BSE$ and calculate the imaginary part of the dielectric function\cite{A. L. Fetter}.  
In our calculations, due to the intrinsic anisotropy of 8-Pmmn borophene, we compute the absorption spectra for incident radiation whose field vector is polarized along two different crystallographic directions.      	    
The zero-temperature optical absorption spectrum of 8-Pmmn borophene for an
in-plane incident light with a polarization along the $y$-axis, i.e. [010] direction, is displayed in Fig.\hspace{0.1 cm}\ref{fig:absorbtion}(a) at three levels of approximation; the $LDA-RPA$ which neglects the inter-electron and e-h correlations, $GW-RPA$ method which includes the inter-electron correlation and the $GW-BSE$ approximation that takes into account both the inter-electron and e–h interactions.
As seen in Fig.\hspace{0.1 cm}\ref{fig:absorbtion}, within the $GW-RPA$, the inter-electron correlation shifts the absorption spectra towards higher energy as compared to the $LDA-RPA$ results. When including the e-h coupling in the calculation, the absorption spectrum is changed, dramatically. In general, in low-dimensional systems due to the strong size confinement and the surrounding vacuum, the screening effect is strongly reduced resulting in strongly bound excitons \cite{W. Wei and T. Jacob}. 	      
From Fig.\hspace{0.1 cm}\ref{fig:absorbtion} we see that the $GW-BSE$ optical absorbance of 8-Pmmn borophene exhibits a global shift towards lower energies as compared to the $GW-RPA$ spectrum and furthermore it exhibits more pronounced excitonic peaks. This indicates strong excitonic effects in the semimetallic phase of borophene as was also obtained for carbon nanotubes which have one-dimensional nature \cite{L.YangM.L.Cohen}, black phosphorus monolayer (phosphorene) due to its quasi-one-dimensional band dispersion \cite{D. Y. Qiu}, graphene \cite{L. Yang} and silicene \cite{W. Wei and} as a result of significant many-body interactions in such nanostructures.
In the case of 8-Pmmn borophene this can be attributed to an effective one-dimensional screening correction due to the screening anisotropy, as well as the parabolic-like band character seen around the $S$ point which is absent in graphene.     	    
The onset of optical absorption of 8-Pmmn borophene along $y$ direction involves an infrared peak around ${0.36}$ eV which is due to interband transitions between states near the Dirac points. Our calculations confirm that low-energy excitation behavior is mainly due to dipole-allowed optical transitions. Actually, in the absence of inversion symmetry, the dipole selection rules are less strict and more transitions are allowed. 
The other three important peaks are marked in Fig.\hspace{0.1 cm}\ref{fig:absorbtion}(a) as ${A}$, ${B}$ and ${C}$. The excitonic peak ${A}$ is observed at $0.94$ eV in the near-infrared region. The corresponding exciton binding energy which is defined as the difference between the excitonic energy obtained from ${GW-BSE}$ and the renormalized single-particle (quasi-particle) energy in the ${GW-RPA}$ spectrum is about $610$ meV for 8-Pmmn borophene along the $y$ direction. This peak is attributed to transitions of the part of Brillouin zone between $Y$ and $\Gamma$ points.

The $B$ and $C$ excitonic peaks, occur at $1.33$ eV and ${1.76}$ eV in the near-infrared and visible region, respectively. The ${B}$ excitonic energy is found to be about ${760}$ meV and the e-h binding energy associated to the ${C}$ peak exhibits a stronger excitonic effect as compared to the $A$ and $B$ cases with a high energy of ${910}$ meV. The excitonic $B$ peak is related to dipole-allowed transitions between the degenerate valence and conduction bands around the ${S}$ point of the Brillouin zone. The more pronounced $C$ peak, as compared to the $A$ one, is due to contributions from the states between $\Gamma$ and $S$. Our results for the excitonic binding energies are comparable with those of 1D \cite{L.YangM.L.Cohen} and 2D \cite{W. Wei and T. Jacob, V. Tran} materials, i.e. we found a binding energy of about $600$ meV.

Because of the anisotropic character of 8-Pmmn borophene, we repeated the calculation for the $x$ crystallographic direction ([100] direction) and show the results in Fig.\hspace{0.1 cm}\ref{fig:absorbtion}(b). 
Notice that the optical absorption spectrum strongly depends on the incident light direction as a consequence of the different band dispersions along the $\Gamma$-$X$ and $\Gamma$-$Y$ directions, owing to the non-symmetric crystal structure along the corresponding crystallographic axes.
The most intense excitonic peak is found around $1.27$ eV, marked as $I$, which our calculations of the dipole oscillator strength and amplitude of the excitons reveal that it is originated from transitions around the $S$ point. The corresponding binding energy for the excitons along this direction is about $620$ meV. 
The difference with the $y$-polarization result can be explained by the fact that the electron-hole interactions are influenced by the different screening effects along $x$ and $y$ directions as a consequence of the screening anisotropy in 8-Pmmn borophene and the anisotropic band structure.
Thus, our results predict that 8-Pmmn borophene is a stronger absorber along $y$ direction which can be understood from the selection rules for optical transitions that depend on the crystal symmetries. 
Actually for the $y$ direction, the transition rules are less stringent due to the lower symmetry so more optical transitions are allowed. 

The observed anisotropic characteristic of the absorption spectra is more apparent in the $GW-BSE$ results which indicates the significance of e-h many-body effects in obtaining an accurate description of 8-Pmmn borophene absorption.  
\begin{figure} [ht]
	\includegraphics [width=0.95\linewidth]{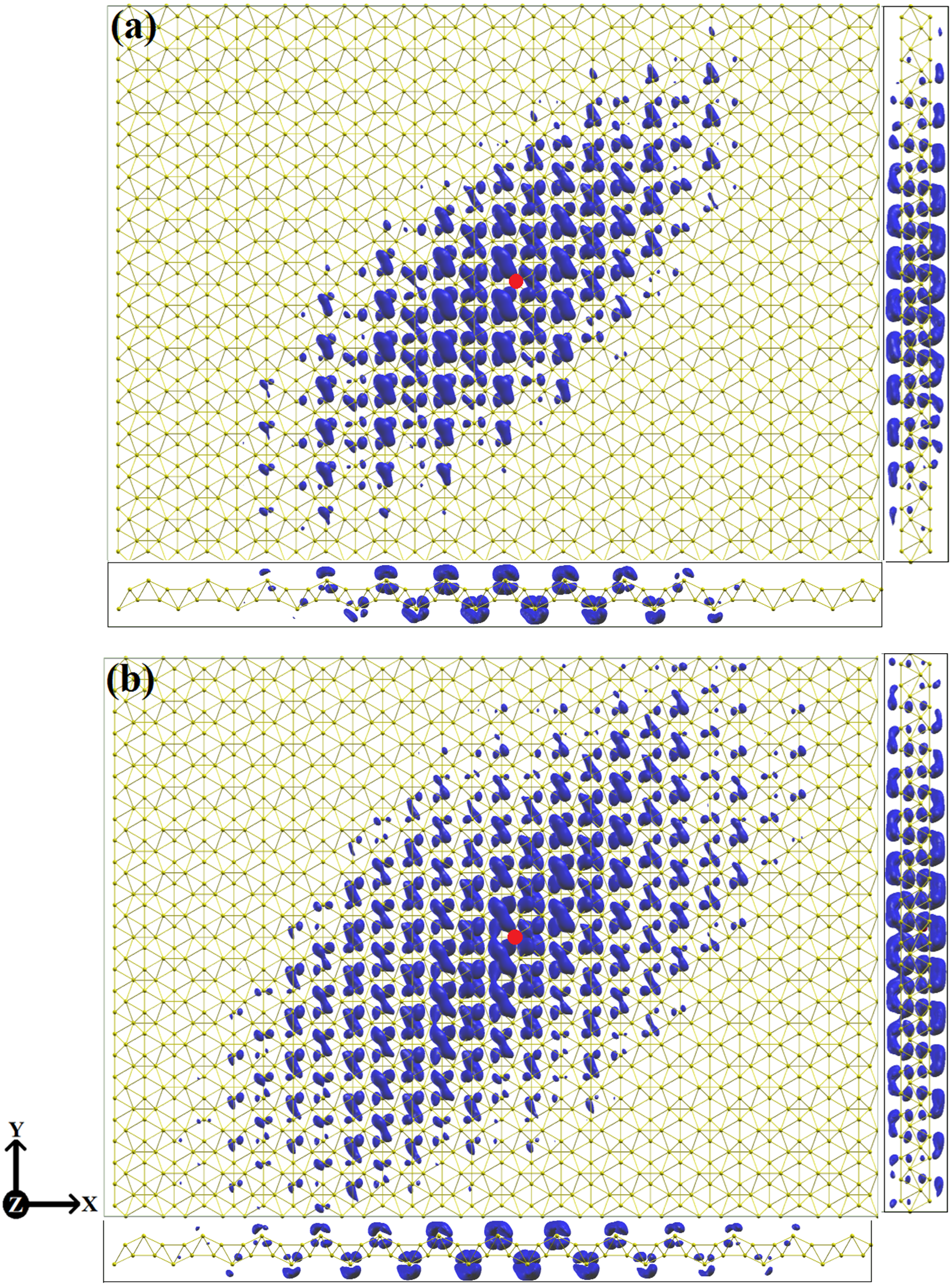}  
	\caption{Excitonic wave function:  top and side views along $y$ (right inset) and $x$ (bottom inset) directions of the three-dimensional electron probability distribution of (a) the exciton $B$ and (b) the exciton $I$. The hole position is fixed and represented by the red dot.}
	\centering
	\label{fig:excitons22}
\end{figure}

\subsection{Excitonic wave functions}

For a more in-depth study of the optical absorption spectra, we show the excitonic wave functions of the bright bound excitons in Fig.\hspace{0.1 cm}\ref{fig:excitons22}. Here, the hole position (the red dot located on the boron atom in the hexagonal sublattice) is fixed and the real-space three-dimensional electron probability distribution is plotted. Figs.\hspace{0.1 cm}\ref{fig:excitons22}(a) and (b) show the top and side views (along $x$ and $y$ directions) of the excitonic wave function corresponding to the bound excitons $B$ (marked in Fig.\hspace{0.1 cm}\ref{fig:absorbtion}(a)) and $I$ (marked in Fig.\hspace{0.1 cm}\ref{fig:absorbtion}(b), respectively. As can be observed, the electron probability densities of both excitons are distributed anisotropically along the $x$ and $y$ directions due to the anisotropic nature of 8-Pmmn borophene crystal along these two directions. Moreover, the anisotropy causes the difference between the binding energies of the excitons (as mentioned in part 3.2, the binding energies of the excitons $B$ and $I$ are 760 meV and 620 meV, respectively) therefore, the wave function of the exciton $I$ is less localized than the exciton $B$. This is clearly shown in Fig.\hspace{0.1 cm}\ref{fig:excitons22} and as expected, the average radius of the exciton I is larger than B. In addition, the electron probability of exciton $I$ has not a significant distribution around the boron atoms of the hexagonal sublattice. The side views of the excitonic wave functions shown in the insets of Fig.\hspace{0.1 cm}\ref{fig:excitons22}, also indicate the strong anisotropy of the exciton wave function. On the other hand, our calculations show that the excitons $A$ and $C$ arising from transitions around the Dirac point, have spatially delocalized wave functions (not shown here).

\section{Conclusions}

The optical properties of 8-Pmmn borophene nanosheet with an anisotropic band structure have been studied by taking into account the many-body e-e and e-h interactions. First, a more reliable band structure was obtained by including the e-e interaction into our calculations through the $GW$ approximation. The quasi-particle energies and the renormalization of the Fermi velocity for 8-Pmmn borophene were obtained. The anisotropic Dirac cone and semimetal characteristics that are predicted from the DFT-LDA, were confirmed and an increased bandgap was obtained at the $S$ and $\Gamma$ symmerty points. The Fermi velocity is enhanced when using the $GW$ approach.   
The optical absorption spectra of 8-Pmmn borophene were calculated using many-body perturbation theory at three levels of approximation: $LDA-RPA$, $GW-RPA$ and $GW-BSE$.  
When taking into account the e-h interaction through the $BSE$ formalism leads to a redshift of the absorbance peaks of 8-Pmmn borophene as compared to the $GW-RPA$ results. It also led to strong absorption peaks due to the formation of excitons. Another interesting optical absorption feature of this phase of borophene is its anisotropic characteristic as a result of the band structure and screening anisotropy. States around the $S$ symmetry point of the Brillouin zone are responsible for the important $B$ and $I$ excitons in the $y$ and $x$ directions, respectively. The highly anisotropic optical absorption properties of 8-Pmmn borophene makes it a very promising material for optoelectronic applications in e.g. polarization-sensitive photodetectors.  

\section*{Conflicts of interest}
There are no conflicts to declare.


\renewcommand\refname{References}




\end{document}